\documentclass[preprint]{elsarticle}
\usepackage{amsmath,amssymb,amsfonts}
\usepackage{algorithm}
\usepackage{algorithmicx}
\usepackage{algpseudocode}
\usepackage{graphicx}
\usepackage{tabularx}
\usepackage{multirow}
\usepackage{multicol}
\usepackage{booktabs}
\usepackage{setspace}
\usepackage{textcomp}
\usepackage{indentfirst}
\usepackage{lineno,hyperref}
\hypersetup{colorlinks=true}
\journal{Information Sciences}
\biboptions{sort&compress}

\begin{document}
\begin{frontmatter}
\title{MAUIL: Multi-level Attribute Embedding for Semi-supervised User Identity Linkage}

\author[address1]{Baiyang Chen}
%\ead{chenby@stu.xhu.edu.cn}
\author[address1]{Xiaoliang Chen\corref{correspauthor}}
\cortext[correspauthor]{Corresponding author}
\ead{chenxl@mail.xhu.edu.cn}

\address[address1]{School of Computer and Software Engineering, Xihua University, Chengdu 610039, P. R. China}

\begin{abstract}
User identity linkage (UIL) across social networks has recently attracted an increasing amount of attention due to its significant research challenges and practical value. Most of the existing methods use a single method to express different types of attribute features. However, the simplex pattern can neither cover the entire set of different attribute features nor capture the higher-level semantic features in the attribute text. This paper establishes a novel semisupervised model, namely the multilevel attribute embedding for semisupervised user identity linkage (MAUIL), to seek the common user identity across social networks. MAUIL includes two components: multilevel attribute embedding and regularized canonical correlation analysis (RCCA)-based linear projection. Specifically, the text attributes for each network are first divided into three types: character-level, word-level, and topic-level attributes. Second, unsupervised approaches are employed to extract the corresponding three types of text attribute features, and user relationships are embedded as a complimentary feature. All the resultant features are combined to form the final representation of each user. Finally, target social networks are projected into a common correlated space by RCCA with the help of a small number of prematched user pairs.
   We demonstrate the superiority of the proposed method over  state-of-the-art methods through extensive experiments on two real-world datasets. All the datasets and codes are publicly available online\footnote{https://github.com/ChenBaiyang/MAUIL}.
	\end{abstract}


\end{frontmatter}
%\linenumbers

\section{Introduction}
Social networks have become increasingly important in daily life. Commonly, people join multiple social networks to enjoy different types of services simultaneously, e.g., Facebook and Twitter. Users usually have separate accounts in different social networks. These accounts can act as bridges connecting the networks. The problem of user identity linkage (UIL), which aims to link the identities of the same natural person across different social platforms, has become increasingly urgent among national security, public opinion supervision, and business recommendation entities worldwide. Developing a highly accurate UIL model contributes greatly to constructing a comprehensive view of user characteristics. An increasing number of applications, including friend recommendation \cite{Shu20175}, cross-network information diffusing prediction \cite{Zafarani2014635}, and link prediction \cite{Zhang20141286}, have recognized the necessity and benefits of UIL.

Early studies addressed the UIL problem by leveraging self-reported user attributes, including user profiles \cite{Vosecky2009360, Liu2013495, Zafarani2009354} (e.g., username, gender, location), user-generated content \cite{Kong2013179} (e.g., tweets, posts, publications) and user behaviors \cite{Zafarani201341}(e.g., tagging, habits). Many of these attribute-based solutions were heuristic, handling particular string patterns or similarity comparisons. For example, a study by Zafarani et al. (2009) \cite{Zafarani2009354} created a set of specific rules based on empirical observations of username string patterns, and employed these rules to discover the corresponding identities across networks. Kong et al. (2013) \cite{Kong2013179} converted each user's posts into weighted bag-of-word vectors, based on which the inner product similarity and the cosine similarity were used to find matched user identity pairs. Such methods are sensitive to the similarity measures or the string patterns of user profiles, and therefore share the following two significant shortcomings:
\begin{enumerate}[(i)]
	\item \textbf{Lack of a general approach for dealing with multiple types of attributes.}\\
 The difference in attribute types creates difficulty in developing general methods. A perfect example can be found in the following three types of user attributes. Usernames often have a large proportion of customized words (e.g., Tom008, Tom{\&}Jerry). Affiliation is usually a phrase that consists of some regular words (e.g., professor, university). User posts may include several paragraphs and high-level semantic features (e.g., post topic, user opinion). Consequently, traditional heuristic approaches can only cover some types, but never all. There is no general way to handle attributes of multiple types.
	\item \textbf{Challenge to mine the hidden high-level semantic connections in texts.}\\
It is necessary to catch the hidden connections between different attributes. An example can be seen in the two words ``teacher" and ``professor" that appeared in someone's affiliation. String alignment rules cannot match them. However, ``teacher" and ``professor" are highly related in semantics. Thus, traditional heuristic approaches will not apply in such a scenario.
	\end{enumerate}

Alternatively, the structural information of social networks can be directly exploited for the UIL problem. For instance, Liu et al. (2016) \cite{Liu20161774} proposed learning a network embedding for linking users by modeling the followership/followeeship of each user. However, the scale of real social networks is massive and changes dynamically. It is not easy to obtain complete structural information. Recent attention has focused on the application of incorporating both user attributes and network structure information. Zhong et al. (2018) \cite{Zhong20185714} suggested incorporating an attribute-based model and a relation-based model into the cotraining framework, making them reinforce each other iteratively. Most of the methods use a single method (e.g., bag-of-words or TF-IDF) to represent all the text attributes, and ignore the apparent differences in text composition or the higher semantic features (e.g., topics).

This paper addresses the UIL problem from a multiview perspective. The basic idea is to recognize the user features on target social networks as different views (feature sets) of the same natural person. The concept of isomorphism states that the same natural person's social identity has a high correlation in different social feature spaces. If we can find mappings from the views to the same feature space and maximize the correlations between their projected features, the users' identities can be identified through existing similarity or distance measures. Following the popular assumption that the correlation between the projections of the two views is linear \cite{Man20161823, Mu20161775, Li2018447}, we consider canonical correlation analysis (CCA), which seeks the projections for different variables so that their correlations are maximized, as the solution for finding such mappings. However, the classical CCA requires that the feature dimensions are smaller than the number of observed samples, thus it fails in semisupervised configurations where we do not have many labeled data. Therefore, we adopt the regularized CCA (RCCA) \cite{Warton2008340} which is well adapted to annotate constrained scenarios for the UIL problem, and propose a novel semisupervised model, namely multilevel attribute embedding for semisupervised user identity linkage (MAUIL). %connects user identities between two social networks with text attributes (hereafter referred to as ``attributed social networks"), which
%这一句影响内容连贯性，attributed的概念放到 problem definition部分介绍。
This model involves two components: multilevel user attribute embedding and RCCA-based projection. First, the text attributes for each network are divided into three types: character-level, word-level, and topic-level attributes. To capture the higher-level semantic features and merge the multiple types of attributes without reliance on any text annotations, three unsupervised methods are employed to extract the corresponding text attribute features. User relationships in social networks are considered an extra user feature type, and all the resultant features are combined to form the final representation of each user. Later, target networks are projected into a common correlated space by RCCA with the help of a small number of prematched user pairs. After the projection, identical users from different networks are located nearby.
In other words, a smaller distance in the common space indicates that the candidates have a more probable chance of being the same natural person. We summarize the main research contributions of this paper as follows:
\begin{enumerate}[(i)]
  \item One of the more significant contributions to emerge from this study is that we recommend a multilevel attribute embedding method to address multiple user attribute types. Our method is proven efficient in capturing various feature types and high-level semantic features in text attributes without relying on any text annotations. Moreover, user attributes and network structures are integrated to construct a more promising representation of social network users. Thus, MAUIL is suitable for the unfortunate situation of missing user attributes or sparse network structures, and has strong robustness.
  \item This paper proposes a novel semisupervised UIL model, MAUIL, which consists of multilevel attribute embeddings and RCCA-based linear projection. Unlike existing methods, MAUIL represents the UIL problem from a multiview perspective and links the projections from social networks to shared feature space by maximizing the correlations between these views.
  \item This study extensively evaluates MAUIL on two real-world datasets of social networks and coauthor networks. The results show the superior performance of the proposed model through a comprehensive comparison with state-of-the-art baseline methods.
	\end{enumerate}

The rest of this paper is organized as follows. Section \ref{sec:Related Work} reviews and summarizes the related works. Section \ref{sec:Definition} formally defines the studied problem. Section \ref{sec:model} details the proposed model. Section \ref{sec:experiment} presents the experimental results. Finally, we conclude this work in Section \ref{sec:conclusion}.

\section{Related Work}
\label{sec:Related Work}
The task of user identity linkage (UIL) \cite{Zafarani2009354} was initially defined as connecting identical user identities across communities. The problem is also known as user alignment, network alignment, or anchor link prediction. Therefore, the social networks that share common users are called aligned (matched) networks in some studies, and the shared users who act as anchors aligning the networks are also referred to as aligned users or matched user pairs.

Existing UIL approaches can be roughly categorized into three groups: supervised, semisupervised, and unsupervised methods. Most of the available literature pertains to supervised methods, which aim to learn a ranking model or a binary classifier to identify user identities \cite{Vosecky2009360,Kong2013179,Zhang20141286,Liu20152005,Man20161823,Mu20161775,Zhang201753,Feng201817540,Zhang2018327,Liu201923595,Qiao20192741,Feng2019459,Li202078,Fu2020105301,Li202185}. For example, Man et al. (2016) \cite{Man20161823} employ network embedding technologies to maintain significant structural regularities of networks by prematched links. Similarly, Mu et al. (2016) \cite{Mu20161775} project user identities of multiple networks into a common latent user space and find linked users by comparing their distances. Zhang et al. (2017) \cite{Zhang201753} incorporate both users' global features and local features, based on which a unified framework UniRank is proposed to recognize users' identities. A seminal study by Feng et al. (2018) \cite{Feng201817540} designs three similarity metrics, based on which a two-stage iterative algorithm CPCC is presented to link identical users. Zhang et al. (2018) \cite{Zhang2018327} adopt a multiview (the character view and the word view) attribute embedding mechanism and propose a graph neural network to directly represent the ego networks of two target users into an embedding space for UIL. A graph attention embedding model is proposed by Liu et al. (2019) \cite{Liu201923595}, which exploits social structures by painting the weighted contribution probabilities between followerships and followeeships. Higher-level feature representations formed the central focus of the study by Qiao et al. (2019) \cite{Qiao20192741}, in which a Siamese neural network is proposed to learn the high-level representation of users' web-browsing behaviors.
Li et al. (2020) \cite{Li202078} use the similarity of k-hop neighbors to present the information redundancy of the network structures, which can be applied to perform user identifications by combining name-based user attributes.
Fu et al. (2020) \cite{Fu2020105301} exploit the higher-order structural properties and alignment-oriented structural consistency to learn a unified graph embedding method called MGGE for UA. A recent study by Li et al. (2021) \cite{Li202185} considered UA as a sequence decision problem and proposed a reinforcement learning model called RLINK to align users from a global perspective.

In all the supervised studies reviewed here, the process of collecting enough aligned users as initial annotations is onerous and has a high labor cost. Hence, some unsupervised approaches \cite{Liu2013495,Riederer2016707,Nie2016107,Zhong20185714,Zhou20181178,Heimann2018117,Li2018447,Zhou20202334,Li20211} are proposed to tackle the UIL problem without the use of any labels.
Nie et al. (2016) \cite{Nie2016107} proposed a dynamic core interest mapping algorithm (DCIM) that combines network structures and user article contents to link identities across networks.
Zhou et al. (2018) \cite{Zhou20181178} worked on the UIL problem and proposed a user identification algorithm, which was complemented by an exploitation of friend relationships in social networks without other prior knowledge. Heimann et al. (2018) \cite{Heimann2018117} showed how to capture a node's structural identity from the degrees of its neighbors and how to detect the attribute identity by using an embedding approach. Li et al. (2018) \cite{Li2018447} found that an earth mover's distance can be utilized to measure the abstract distances between the user identity distributions of social networks.
Zhou et al. (2020) \cite{Zhou20202334} proposed capturing node distributions in Wasserstein space and reformulating the UA task as an optimal network transport problem in a fully unsupervised manner. Recently, Li et al. (2021) \cite{Li20211} exploited users' check-in records and considered spatial-temporal information (e.g., location and time) jointly in users' activities to match identical user accounts without any annotations.

Unsupervised approaches are convenient because it is possible to avoid relying on labeled data. However, the very technologies have also created some potential problems in almost all their applications. Unsupervised methods usually suffer from lower performances than supervised methods. Therefore, several semisupervised approaches \cite{Tan2014159,Zhang20151485,Liu20161774,Zhou2016411,Zhao20185834, Li2019996,Li2019249,Zhou20191360,Liu20201824,Liu202036} have recently emerged to solve the UIL problem. The typical characteristic of a semi-supervised approach is the common use of unlabeled samples along with a few annotations. Additionally, unlabeled identity pairs can be expected during the learning process.

Following the above idea, Zhang et al. (2015) \cite{Zhang20151485} proposed an energy-based model (COSNET), which jointly considered the local and global consistency among multiple networks, to increase the accuracy of UIL. Zhou et al. (2016) \cite{Zhou2016411} discovered that user identities could be iteratively linked by calculating network structural matching degrees. Zhang et al. (2018) \cite{Zhao20185834} proposed a unified hypergraph learning framework (UMAH), which intrinsically performs semisupervised manifold alignments using profile information for calibration. Li et al. (2019) \cite{Li2019996} treated all users in a social network as a whole and designed a weakly supervised adversarial learning framework, SNNA, to align users from the distribution level. Later, an extension model, MSUIL \cite{Li2019249}, based on SNNA was presented to address the challenges in multiple social network scenarios. Zhou et al. (2019) \cite{Zhou20191360} focused on model interpretability and propose a semisupervised dNAME model, which embeds nodes in a disentangled manner by tracing the importance of each anchor node and its explanations of the UIL performance. Liu et al. (2020) \cite{Liu20201824} proposed an embedding-based approach, which represents users' follower/followee relationships as input/output context vectors, to recognize users' identities with the help of a few annotations.
Another recent work by Liu et al. (2020) \cite{Liu202036} studied the mutual promotion effect of users and employers in heterogeneous social networks, and they proposed a matrix factorization-based representation learning framework MFRep for network alignments. Unfortunately, the above methods share the same weakness; they can neither cover the entire set of different attribute features nor capture the higher-level semantic features in the attribute text.

\section{Problem Definition}
\label{sec:Definition}
This section examines the basics of the problem. Table \ref{Tab:Notations} summarizes the notations used in this paper.

A social network with text attributes (hereafter referred to as ``attributed social networks") is a three-tuple $G=(V,E,A)$, where $V=\{v_1,v_2,\ldots,v_n\}$ and $E=\{e_{ij}=(v_i,v_j)|v_i,v_j\in V\}$ represent the set of $n$ users and the set of undirected edges between users, respectively. Edge $e_{ij}$ denotes the binary status of the link between user $v_i$ and $v_j$, e.g., friendships on Facebook. $A=(A_c,A_w,A_t)$ is a set of user attributes, e.g., names, affiliations, and education experiences, which consists of three subsets:(1) character-level attributes $A_c$, (2) word-level attributes $A_w$, and (3) topic-level attributes $A_t$.

\begin{table}[h!]
    \caption{Main notations used in this paper}\label{Tab:Notations}
    \centering
    %\begin{spacing}{1.3}
    \begin{small}
    \setlength{\tabcolsep}{3pt}
    \renewcommand{\arraystretch}{1.2}
    \begin{tabular}{p{60pt}<{\centering}|p{200pt}}
    \toprule[1.0pt]
    \textbf{Notation}&\textbf{Description}\\
    \hline
    $G^X/G^Y$   &   Networks $X$ and $Y$ that participate in UIL. \\
    $V$         &   Set of users in a social network.\\
    $E$         &   Set of edges in a social network.\\
    $A$         &   Set of user attributes in a social network.\\
    $\mathbb{R}$ &  Set of real numbers.\\
    $A_c,P_c$   &   Set of character-level attributes and its feature matrix.\\
    $A_w,P_w$   &   Set of word-level attributes and its feature matrix.\\
    $A_t,P_t$   &   Set of topic-level attributes and its feature matrix.\\
    $P_s$       &   Structure feature matrix of a social network.\\
    $d_c,d_w,d_t,d_s$ & Dimensions of character-level, word-level, topic-level, and structure embedding.\\
    $X/Y$       &   Combined feature matrices of $G^X/G^Y$.\\
    $H/M$       &   Canonical matrices of a pair of networks.\\
    $\mathbf{h}_i/\mathbf{m}_i$	& $i$-th pair of canonical vectors.\\
    $C_{XY}$    &	Covariance matrix of $X$ and $Y$.\\
    $\hat{C}_{XX}$&	Regularized covariance matrix of network $X$.\\
    $r^X/r^Y$	&	Regularization coefficients of $C_{XX}/C_{YY}$\\
    $Z^X/Z^Y$   &	Projected feature matrices of $X/Y$.\\
    \bottomrule[1.0pt]
    \end{tabular}
    \end{small}
    %\end{spacing}
	\end{table}

This paper considers this type of scenario: Some users in two different social networks are the same natural person who form the anchor link across the networks. We refer to one of the networks as $G^X=(V^X,E^X,A^X)$ and the other as $G^Y=(V^Y,E^Y,A^Y)$.
This investigation intends to identify, if any, the counterpart in $G^Y$ (resp. $G^X$) for each node in $G^X$ (resp. $G^Y$). The user identity linkage problem can be formalized as follows.

\newdefinition{definition}{Definition}
\begin{definition}
\textbf{User Identity Linkage (UIL)}: Given two attributed social networks $G^X$ and $G^Y$ and a few prematched user pairs $S=\{(v_i,v_j)|v_i\in V^X,v_j\in V^Y \}$, the UIL problem involves locating the other hidden matched identity pairs $L=\{(v_i,v_j)|v_i\in V^X,v_j\in V^Y,(v_i,v_j)\notin S\}$, where $v_i$ and $v_j$  belong to the same natural person.
\end{definition}

This study proposes a semisupervised model, MAUIL, to solve the UIL problem. In this model, the networks $G^X$ and $G^Y$ are independently embedded into two latent feature spaces $P^X\in \mathbb{R}^{d\times n}$ and $P^Y\in \mathbb{R}^{d\times n}$, where $d$ and $n$ denote the number of dimensions and the number of users, respectively. In addition, $G^X$ and $G^Y$ are mapped into a common correlated space by using two projections $H\in \mathbb{R}^{d\times k}$ and $M\in \mathbb{R}^{d\times k}$, respectively, where superscript $k$ represents the number of projection vectors.

\section{Model Specification}
\label{sec:model}
\subsection{Overview}

\begin{figure*}[!h]
    \begin{center}
    \includegraphics[width=\textwidth]{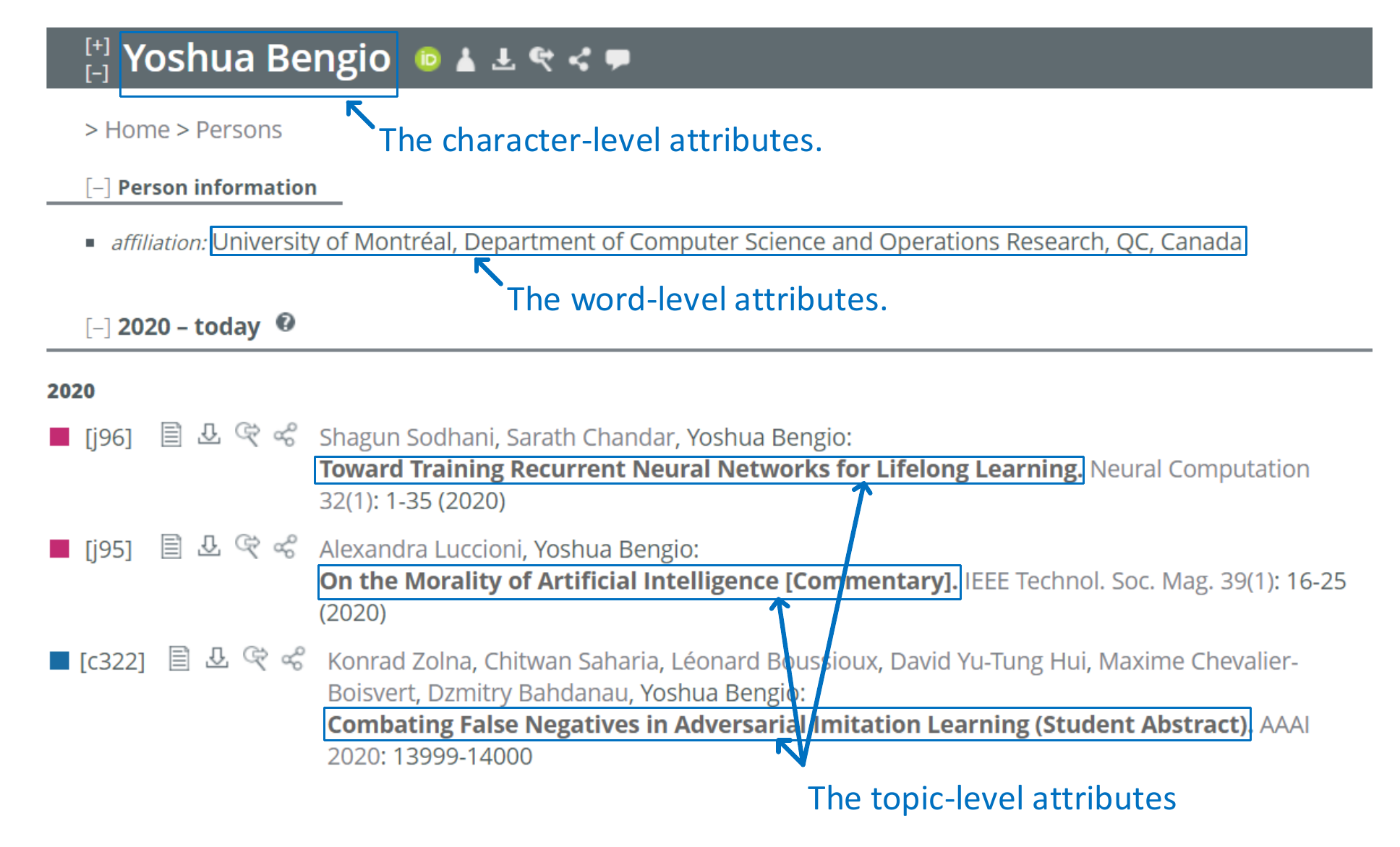}
    \caption{\small A case of text attributes.}\label{Fig:case_textattrubutes}
    \end{center}
    \end{figure*}

This section presents the details of MAUIL. The model has two components: multilevel attribute embedding and RCCA-based projection. We utilize MAUIL to detect the potential user identity links covering the target networks $G^X$ and $G^Y$. First, the text attributes of each network are divided into three types: character-level attributes $A_c$, word-level attributes $A_w$, and topic-level attributes $A_t$. An example of the text attributes can be seen in Figure \ref{Fig:case_textattrubutes}. Second, three unsupervised methods (discussed in Sections \ref{sec:char}, \ref{sec:word}, and \ref{sec:topic}) are employed to generate three corresponding feature matrices $P_c$, $P_w$, and $P_t$, respectively. In addition, the users’ relationships in the social networks has been proven to be beneficial for UIL tasks. In this study, user relationships evolved as a particular type of user feature $P_s$ (discussed in Section \ref{sec:network}).
As a result, a total of four feature matrices $P_c$, $P_w$, $P_t$, and $P_s$ are combined to form the final representation of the target social networks $G^X$ and $G^Y$.

RCCA is usually used to explore the correlation between two multivariables (vectors). In this study, we introduce RCCA to find the mappings that can project the networks $G^X$ and $G^Y$ into a common correlated space (discussed in Section \ref{sec:CCA}).

\begin{figure*}[!h]
    \begin{center}
    \includegraphics[width=\textwidth]{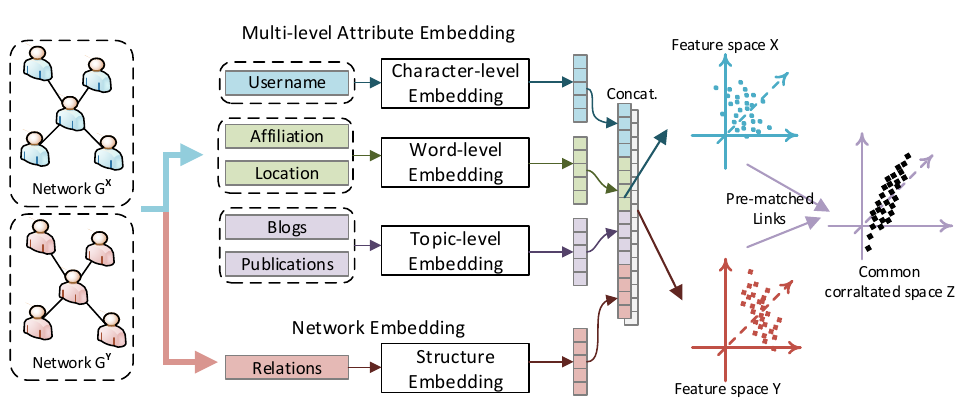}
    \caption{The overview of MAUIL model, including two components: multi-level attribute embedding and RCCA based projection.}\label{Fig:MAUIL}
    \end{center}
    \end{figure*}

For any user belonging to network $G^X$ or $G^Y$, we identify the most likely counterpart in the other network by comparing their distances in the common correlation space. A closer distance indicates that the two users are more likely to be the same natural person. An overview of MAUIL is presented in Figure \ref{Fig:MAUIL}.

\subsection{Multi-level Attribute Embedding}
\label{sec:emb}
In this section, different social networks are embedded in the same way. Hence, we use the same notation $G$ without distinguishing between $G^X$ and $G^Y$.
\subsubsection{Character-level attribute embedding}
\label{sec:char}
This section intends to preserve the text similarity of similar users through character-level attribute embedding.

The character-level attribute of user $v_i$ in a social network $G$ with $n$ users is denoted by $a_{i}^c$, which is specified as the username string that may contain part of the characters, spaces, and special symbols. Attribute $a_{i}^c$ can be divided into a series of $m$ unique tokens $\mathbf{w}={w_1,w_2,\ldots,w_k,\ldots,w_m}$, including characters, numbers, and q-grams. The vector form of $a_{i}^c$ refers to a count-weighted expression $\overrightarrow{\mathbf{x}_{i}^c}=[\mathbf{x}_{w_1},\mathbf{x}_{w_2},\ldots,\mathbf{x}_{w_k},\ldots,\mathbf{x}_{w_m}]\in \mathbb{R}^{m\times 1}$, where for any $k\in\{1,2,\ldots,m\}$, $\mathbf{x}_{w_k}$ is the count number of the corresponding piece $w_k$ in $\mathbf{w}$.

This study constructs the count-weighted expression by performing a simple bag-of-word vectorization calculation.
The above formalizations can be illustrated by user $v_i$ shown in Figure \ref{Fig:case_textattrubutes}. In terms of the username, we have the character-level attribute $a_{i}^c=$``Yoshua Bengio'', which can be divided into a series of tokens $\mathbf{w}=$`a':1,`b':1,`c':0,$\ldots$,`o':2,$\ldots$, others:0.
Their count numbers follow each token in the text. Finally, the corresponding count-weighted vector is
$\overrightarrow{\mathbf{x}_{i}^c}=[1,1,0,\ldots,2,\ldots]$.

Having discussed how to represent the character-level attribute of a user, the following descriptions address ways of considering all the $n$ users in a social network. The notation $A_c=\{a_{1}^c,a_{2}^c,\ldots,a_{i}^c,\ldots,a_{n}^c\}$ is used here to refer to the set of character-level attributes for $n$ users. Similarly, the vectorization of set $A_c$ can be represented as the count-weighted vector sequences $X_c=[\overrightarrow{\mathbf{x}_{1}^c},\overrightarrow{\mathbf{x}_{2}^c},\ldots,\overrightarrow{\mathbf{x}_{i}^c},\ldots,\overrightarrow{\mathbf{x}_{n}^c}]^{T}$
that are also called the character-level count-weighted matrix.

Data dimensionality reduction technologies were initially applied in deep image learning, and have gradually evolved for use in the NLP area, achieving excellent results. Autoencoders are a necessary and effective way to reduce the dimensionality of data and lessen the computational the burden of deep learning. This study employs a one-layer autoencoder to embed the count-weighted vectors into distributed representations. First, the autoencoder maps an input vector $\overrightarrow{\mathbf{x}}$ to a hidden representation $\overrightarrow{\mathbf{z}}$ through a deterministic function $\overrightarrow{\mathbf{z}}=f(\overrightarrow{\mathbf{x}})=W\overrightarrow{\mathbf{x}}+b$, where $W\in \mathbb{R}^{d_c\times m}$ and $b\in \mathbb{R}^{d_c\times 1}$  are the weight matrix and a bias vector, respectively. Second, the resulting latent representation $\overrightarrow{\mathbf{z}}$ is mapped back to a ``reconstructed'' vector $\overrightarrow{\mathbf{y}}$ in input space $\overrightarrow{\mathbf{y}} = g (\overrightarrow{\mathbf{z}})= W^*\overrightarrow{\mathbf{z}} + b^*$, where $W^*\in \mathbb{R}^{d_c\times m}$ is the weight matrix of the reverse mapping function $g(.)$.

This study optimizes the parameters to minimize the average reconfiguration loss with:
\begin{equation}\label{AE_loss}
	L=\frac{1}{n}\sum\|x_i-y_i\|^2
	\end{equation}
After optimization, we obtain the character-level feature matrix $P_c=[\overrightarrow{\mathbf{p}_{1}^c},$ $\overrightarrow{\mathbf{p}_{2}^c},\ldots,\overrightarrow{\mathbf{p}_{i}^c},\ldots,\overrightarrow{\mathbf{p}_{n}^c}]^{T}$
of the character-level attribute $A_c$ with $d_c$ dimensions by:
\begin{equation}\label{char_emb}
	P_c= WX_c+b
	\end{equation}
where $W$ and $b$ are the weight matrix and the bias of the autoencoder. For any $i\in\{1,2,\ldots,n\}$, $\overrightarrow{\mathbf{p}_{i}^c}$ is the character-level feature vector of the count-weighted vector $\overrightarrow{\mathbf{x}_{i}^c}$. Algorithm \ref{alg:char_emb} presents the entire procedure of character-level attribute embedding with a function called CharLevelEmb.
\renewcommand{\algorithmicrequire}{ \textbf{Input:}}
\renewcommand{\algorithmicensure}{ \textbf{Output:}}
\begin{algorithm}[h!]
    \caption{Character-level attribute embedding.}
    \label{alg:char_emb}
    \begin{small}
    \begin{algorithmic}[1]
    \Require The character-level attribute set $A_c$ and the character-level embedding dimension $d_c$.
    \Ensure The character-level attribute feature matrix $P_c$.
    \Function{CharLevelEmb}{$A_c,d_c$}
        \State Initialize $X_c$
        \For {$\forall a_{i}^c$ in $A_c$}
            \State $\mathbf{w}=CharTokenize(a_{i}^c)$
            \State $\overrightarrow{\mathbf{x}_{i}^c}=CountVector(\mathbf{w})$
            \State $X_c.append(\overrightarrow{\mathbf{x}_{i}^c})$
        \EndFor
        \State Initialize $W,b$
        \Repeat
            \For {$\forall \overrightarrow{\mathbf{x}}$ in $X_c$}
                \State $\overrightarrow{\mathbf{z}}=f(\overrightarrow{\mathbf{x}})=W\overrightarrow{\mathbf{x}}+b$
                \State $\overrightarrow{\mathbf{y}}=g(\overrightarrow{\mathbf{z}})=W\overrightarrow{\mathbf{z}}+b$
                \State update $W,b$ based on Eq. $\ref{AE_loss}$
            \EndFor
        \Until Convergence
        \State $P_c \gets$ Eq. $\ref{char_emb}$
        \State \textbf{return} $P_c$
    \EndFunction
    \end{algorithmic}
    \end{small}
	\end{algorithm}

\subsubsection{Word-level attribute embedding}
\label{sec:word}
Word2vec \cite{Mikolov20131301,Guo2021131} is one of the most common procedures for transforming texts to feature vectors. This section applies the Word2vec approach to capture the word-level characteristics of user attributes.

The word-level attribute of user $v_i$ in a social network is represented as $a_{i}^w$, which consists of phrases or short sentences such as gender, location, affiliations, and education experiences. The attribute $a_{i}^w$ can be split into a sequence of total $m$ words $\mathbf{w}_i=w_{i1},w_{i2},\ldots,w_{ik},\ldots,w_{im}$, where $w_{ik}$ is the vocabulary expression of the $k$-th word in $a_{i}^w$. A more detailed description of the above formalization is given in the example of the user ``Yoshua Bengio'', which is shown in Figure \ref{Fig:case_textattrubutes}. His word-level attribute is $a_{i}^w=$``University of Montréal, Department of Computer Science and Operations Research, QC, Canada'', which is tokenized into a sequence of words followed by its count number as follows:
$\mathbf{w}_i=$``University'':1,``of'':2,``Montréal'':1, ``Department'':1, $\ldots$,others:0. Each word is later represented by a unique number as its ID in the vocabulary.

We use $A_w=\{a_{1}^w,a_{2}^w,\ldots,a_{i}^w,\ldots,a_{n}^w\}$ to denote the set of word-level attributes in a social network with $n$ users. Each $\mathbf{w}_i$ of the attribute $a_{i}^w$ is treated as a word-level document for deep learning. Then, all the $n$ documents constitute a corpus. We plan to train the word vectors in the corpus by using the language model CBOW \cite{Mikolov20131301}. In the training process, for any $k\in\{1,2,\ldots,m\}$, the word vector of $w_{ik}$ in document $\mathbf{w}_i$ is a $d_w$-dimensional vector $\overrightarrow{\mathbf{x}_{ik}}\in R^{d_w\times 1}$. Hence, the word-level attribute vector of user $v_i$ can be denoted as $\overrightarrow{\mathbf{p}_i^w}$ and is calculated by summing the word vectors of all the words in document $\mathbf{w}_i$:\\
\begin{equation}\label{word_emb_sum}
	\overrightarrow{\mathbf{p}_i^w}=\sum_{w_{ik}\in \mathbf{w}_i} \overrightarrow{\mathbf{x}_{ik}}
	\end{equation}
There is a particular problem with the practice of word-level attributes. Word-level attributes compared with character-level attributes may be missing or not distinguishable enough to provide sufficient evidence. The solution is to consider the additional evidence offered by their neighbors according to the homophily principle \cite{Beukel2019361}. Thus, we smooth the embedding of each user by their neighbor embeddings that are regularized by a real number parameter $\lambda\in [0,1]$:\\
\begin{equation}\label{word_emb_ave}
	\overrightarrow{\mathbf{p}_i^w}=(1-\lambda) \sum_{w_{ik}\in \mathbf{w}_i}{\overrightarrow{\mathbf{x}_{ik}}} +\lambda \frac{1}{s_i} \sum_{j\in \mathcal{N}_{i}}\sum_{w_{jk}\in \mathbf{w}_j}{\overrightarrow{\mathbf{x}_{jk}}}
	\end{equation}
where $\mathcal{N}_{i}=\{v_j|(v_i,v_j)\in E\}$ is the neighbor set of  user $v_i$ and $s_i=|\mathcal{N}_{i}|$ is the number of neighbors.
As a result, the word-level attributes $A_w$ of all users can be represented by a word-level feature matrix $P_w=[\overrightarrow{\mathbf{p}_{1}^w},\overrightarrow{\mathbf{p}_{2}^w},\ldots,\overrightarrow{\mathbf{p}_{i}^w},\ldots,\overrightarrow{\mathbf{p}_{n}^w}]^T$. Algorithm \ref{alg:word_emb} presents the detailed procedure of word-level attribute embedding by a function, called WordLevelEmb.

\begin{algorithm}[h!]
    \caption{Word-level attribute embedding.}
    \label{alg:word_emb}
    \begin{small}
    \begin{algorithmic}[1]
    \Require The word-level attribute set $A_w$, the word-level embedding dimension $d_w$, the regularization real number parameter $\lambda$, and the network edge set $E$.
    \Ensure The word-level attribute feature matrix $P_w$.
    \Function{WordLevelEmb}{$A_w,d_w,\lambda,E$}
        \State $wordvectors = Word2Vec(A_w,dim=d_w)$
        \State Initialize $P_w \gets \mathbf{0}$.
        \For {$\forall a_{i}^w$ in $A_w$}
            \State $\mathbf{w}_i=WordTokenize(a_{i}^w)$
            \State Initialize $\overrightarrow{\mathbf{p}_{i}^w} \gets \mathbf{0}$.
            \For {$\forall w_{ik}$ in $\mathbf{w}_i$}
                \State $\overrightarrow{\mathbf{x}_{ik}}=LookUpEmbeddings(wordvectors,w_{ik})$
                \State $\overrightarrow{\mathbf{p}_{i}^w} += \overrightarrow{\mathbf{x}_{ik}}$   (Eq. $\ref{word_emb_sum}$)
                \State $P_w.append(\overrightarrow{\mathbf{p}_{i}^w})$
            \EndFor
        \EndFor
        \State $P^*_w = copy(P_w)$
        \For {$\forall a_{i}^w$ in $A_w$}
            \State Initialize $(\overrightarrow{\mathbf{p}_{i}^{w}})^* \gets \mathbf{0}$.
            \State $\mathcal{N}_i =GetNeighbors(E,v_i)$
            \For {$\forall v_j$ in $\mathcal{N}_i$}
                \State $\overrightarrow{\mathbf{p}_{j}^w} = P^*_w[j]$
                \State $(\overrightarrow{\mathbf{p}_{i}^{w}})^* += \overrightarrow{\mathbf{p}_{j}^w}$
            \EndFor
            \State $\overrightarrow{\mathbf{p}_{i}^w} = (1-\lambda)\overrightarrow{\mathbf{p}_{i}^w}+\frac{\lambda}{s_i}(\overrightarrow{\mathbf{p}_{i}^w})^*$ (Eq. $\ref{word_emb_ave}$)
            \State Update $P_w \gets \overrightarrow{\mathbf{p}_{i}^w}$.
        \EndFor
        \State \textbf{return} $P_w$
    \EndFunction
    \end{algorithmic}
    \end{small}
	\end{algorithm}

\subsubsection{Topic-level attribute embedding}
\label{sec:topic}
This study employs latent Dirichlet allocation (LDA) \cite{Kim201915,Wan2020243} to extract the topic-level features of user attributes.

The attribute texts of user $v_i$ in a social network, consisting of paragraphs or articles such as posts, blogs, and publications, are combined to form the topic-level attribute $a_{i}^t$. In a similar case shown in Figure \ref{Fig:case_textattrubutes}, the texts of user ``Yoshua Bengio'' include three publications. Their titles or full texts are combined to form the $a_{i}^t$. Hence, we have the set of topic-level attributes $A_t=\{a_{1}^t,a_{2}^t,\ldots,a_{i}^t,\ldots,a_{n}^t\}$ for all users in the social network DBLP, and the $i$-th topic-level attribute $a_{i}^t$ in $A_t$ is treated as the topic-level document $\mathbf{w}_i$.

LDA is a kind of document-topic generation model. A document can contain multiple topics, and each word in the document is generated by one of the topics. LDA can present the topic of each document in a document set as a probability distribution. For example, a document $\mathbf{w}_i$ can be modeled by a topic distribution $\theta_{\mathbf{w}_i}$. As a result, LDA extracts a few words in the vocabulary to describe each topic. In this study, we consider the topic distributions as the topic-level features of user attributes for the UIL problem.

Let $\mathbf{z}=\{1,2,\ldots,d_t\}$ be a set of topic indices. Notation $z_i\in \mathbf{z}$ refers to the topic index of document $\mathbf{w}_i$. A latent topic with index $z_i$ in LDA is characterized by exactly the word distribution $\phi_{z_i}$. Topic distributions provide concise representations of documents. The LDA process is as follows:
\\
    \par\setlength{\parindent}{1em}1. For each topic index $z\in \mathbf{z}$:
    \par\setlength{\parindent}{2em}(a) Choose a word distribution $\phi_z\sim Dirichlet(\beta)$.
    \par\setlength{\parindent}{1em}2. For each document $\mathbf{w}_i$ in a corpus:
    \par\setlength{\parindent}{2em}(a) Choose a topic distribution $\theta_{\mathbf{w}_i} \sim Dirichlet(\alpha)$.
    \par\setlength{\parindent}{2em}(b) For each word $w_i$ of the $\overline{\mathcal{N}}$ words in the model vocabulary:
    \par\setlength{\parindent}{3em}a. Choose a topic with the index $z_i\in \mathbf{z}\sim Multinomial(\theta_{\mathbf{w}_i})$.
    \par\setlength{\parindent}{3em}b. Choose a word $w_i\sim Multinomial(\phi_{z_i})$.\\
    \setlength{\parindent}{0.8em}
\\
where $Dirichlet(\cdot)$ and $Multinomial(\cdot)$ denote the Dirichlet distribution and multinomial distribution, respectively. $\alpha$ and $\beta$ are hyperparameters of the Dirichlet distribution, specifying the property of the priors on the topic and word distributions. $\overline{\mathcal{N}}$ represents the total number of words in the vocabulary used in the model.

In general, let $\mathbf{w}$ be a document, $\mathbf{z}$ be a set of topic indices that identify topics, $\theta$ be a topic distribution, $\phi$ be a word distribution, and $\phi_z$ be a word distribution with respect to topic index $z_i\in\mathbf{z}$. LDA results in the following joint distribution:
\begin{equation}\label{LDA_p}
	p(\mathbf{w},\mathbf{z},\theta,\phi|\alpha,\beta) =p(\phi|\beta)p(\theta|\alpha)p(\mathbf{z}|\theta)p(\mathbf{w}|\phi_z)
	\end{equation}
Then, the posterior distribution $p(\mathbf{z|\mathbf{w}})$ can be calculated by a Gibbs sampling as:\\
\begin{equation}\label{LDA_Gibbs}
    \begin{split}
    p(z_i=j|\mathbf{z}_{-i},\mathbf{w})&\propto \frac{WT_{-i,j}^{(w_i) }+\beta}{\sum_{i=1}^{N}{WT_{-i,j}^{(w_i)}} +\overline{\mathcal{N}}\beta}\\
    &\quad \cdot \frac{DT_{-i,j}^{(\mathbf{w})}+\alpha}{\sum_{k=1}^{d_t}{DT_{-i,k}^{(\mathbf{w})} +d_t \alpha}}
    \end{split}
    \end{equation}
where $z_i=j$ represents the topic that generates the word $w_i$, which is just being indexed by $j\in\mathbf{z}$. Notation $\mathbf{z}_{-i}$ denotes all other topics except for $z_i$. In a corpus, i.e., a set of documents, a word may appear more than once in different documents.
Hence, the count relation between words and topics can be represented as a matrix $WT$. An entry of $WT$, denoted by $WT_j^{(w_i)}$, means the number of words $w_i$ assigned to topic $j$. Similarly, the number of words $w_i$ assigned to topic $j$ that appeared in a corpus without including the current $w_i$ itself is depicted by the notation $WT_{-i,j}^{(w_i)}$. Correspondingly, we have the count relation matrix $DT$ between the documents and topics. An entry of $DT$, denoted by $DT_j^{(\mathbf{w})}$, means the total number of all words in document $\mathbf{w}$ assigned to topic $j$. Similarly, the total number of all words in document $\mathbf{w}$, which have been assigned to topic $j$ without including the current $w_i$ itself, is depicted by the notation $DT_{-i,j}^{(\mathbf{w})}$.

For a numbered document $\mathbf{w}_i$, the estimate of a topic distribution $\theta_{\mathbf{w}_i}$ over topic $j$ can be calculated as:
\begin{equation}\label{LDA_theta}
    \hat{\theta}_j^{(\mathbf{w}_i)}=\frac{DT_j^{\mathbf{w}_i}+\alpha}{\sum_{k=1}^{d_t}{DT_k^{(\mathbf{w}_i)}+d_t\alpha}}
    \end{equation}
Thus, we can obtain the topic probability vector of each user $v_i$ as $\overrightarrow{\mathbf{p}_{i}^t}=\{[\hat{\theta}_j^{(\mathbf{w}_i}])\}_{j=1}^{d_t}$, which is the proposed topic-level attribute feature vector. Consequently, the topic-level attributes $A_t$ of all the users can be represented by a topic-level feature matrix $P_t=[\overrightarrow{\mathbf{p}_{1}^t},\overrightarrow{\mathbf{p}_{2}^t},\ldots,\overrightarrow{\mathbf{p}_{i}^t},\ldots,\overrightarrow{\mathbf{p}_{n}^t}]^T$.
The pseudocode function, called TopicLevelEmb, of the topic-level attribute embedding is shown in Algorithm \ref{alg:topic_emb}.

\begin{algorithm}[h!]
    \caption{Topic-level attribute embedding.}
    \label{alg:topic_emb}
    \begin{small}
    \begin{algorithmic}[1]
    \Require The topic-level attribute set $A_t$, the topic-level embedding dimension $d_t$, and the hyper-parameters $\alpha$ and $\beta$ of Dirichlet distribution.
    \Ensure The topic-level attribute feature matrix $P_t$.
    \Function{TopicLevelEmb}{$A_t,d_t,\alpha,\beta$}
        \State Initialize topic assignments $\mathbf{z}$, and counters $DT, TW$
        \For{each iteration}
            \State update $\mathbf{z},DT,TW$ based on Gibbs Sampling Eq. $\ref{LDA_Gibbs}$
        \EndFor

        \State Initialize $P_t\gets \mathbf{0}$.
        \For{$\forall\mathbf{w}_i$ in $A_t$}
            \State Initialize $\overrightarrow{\mathbf{p}_{i}^t}\gets \mathbf{0}$.
            \For {$j=1:d_t$}
                \State $ \hat{\theta}_j^{(\mathbf{w}_i)} \gets$ Eq. $\ref{LDA_theta}$
                \State $\overrightarrow{\mathbf{p}_{i}^t}.append( \hat{\theta}_j^{(\mathbf{w}_i)})$
            \EndFor
            \State $P_t.append(\overrightarrow{\mathbf{p}_{i}^t})$
            \EndFor
        \State \textbf{return} $P_t$
    \EndFunction
    \end{algorithmic}
    \end{small}
 \end{algorithm}

\subsubsection{Network Structure Embedding and Feature Combining}
\label{sec:network}
The use of network structure embedding has been well accepted for its distinct advantages in the UIL problem. Network structures have played an increasingly important role in helping researchers find similar user relations. The core reason for utilizing structure information is to map the network into a latent space such that users with similar structural roles are positioned close to each other. Several successful methods currently exist for network embedding. This study empirically chooses the LINE method \cite{Tang20151067} to embed the network into the feature matrix $P_s$ with $d_s$ dimensions.

Consequently, we combine the three developed attribute features $P_c$, $P_w$, and $P_t$ with the structural features $P_s$ by a concatenate operation to obtain the final representation of the social network $G^X/G^Y$, denoted as:
\begin{equation}\label{NE_comb}
    \begin{split}
    X&=[P_c^X;P_w^X;P_t^X;P_s^X ]\in \mathbb{R}^{d\times n}\\
    Y&=[P_c^Y;P_w^Y;P_t^Y;P_s^Y ]\in \mathbb{R}^{d\times n}
    \end{split}
    \end{equation}
where $d=d_c+d_w+d_t+d_s$ is the dimension of the combined feature matrix. To bring these different features onto the same scale, we further impose standardization on the feature matrices by centering each row at mean 0 with standard deviation 1. The standardization of feature matrices can establish different features on the same scale. In this paper, each row of a matrices’ standardization center is set to mean 0 with standard deviation 1.

\subsection{RCCA based projection}
\label{sec:CCA}
Section \ref{sec:emb} showed how to establish the combined feature metrics $X$ and $Y$ using the text attributes and the structures of a prematched network pair $G^X/G^Y$. This section examines a method to implement user identity links by proposing two feature metrics. First, we follow the widespread assumption \cite{Mu20161775, Li2018447} that the correlations between the projections of linked networks are linear. In addition, regularized canonical correlation analysis (RCCA) \cite{Hardoon20042639} is particularly useful in maximizing their correlations. Hence, this study adopts an RCCA approach to investigate the identity links between similar users for the UIL problem.

RCCA approaches mostly define the canonical matrices as $H=[\mathbf{h} _1,\mathbf{h} _2,\ldots,$ $\mathbf{h}_i,\ldots,\mathbf{h}_k]\in \mathbb{R}^{d\times k}$ and $M=[\mathbf{m} _1,\mathbf{m} _2,\ldots,$ $\mathbf{m}_j,\ldots,\mathbf{m}_k]\in \mathbb{R}^{d\times k}$, including
$k$ pairs of linear projections. The canonical matrices of the UIL problem are solved by mapping the vector reputation $X/Y$ of the associated social networks $G^X/G^Y$ into a common correlated space $Z$ through a series of well-designed linear projections to maximize the correlation $\rho$ between $H^TX$ and $M^TY$. Then, the potential user identity links between $G^X$ and $G^Y$ can be estimated by comparing the distance of their vectorization features in $Z$. The detailed steps of an RCCA-based projection are specified below.

First, for each pair of canonical vectors $\mathbf{h}_i\in \mathbb{R}^{d\times 1}$ of $X$ and $\mathbf{m}_j\in \mathbb{R}^{d\times 1}$ of $Y$, the purpose of RCCA is to maximize the correlation $\rho$ between $\mathbf{h}_i^TX$ and $\mathbf{m}_j^TY$ \cite{Hardoon20042639}, i.e.,:
\begin{equation}\label{rho}
    \begin{split}
       \rho &=\max corr(\mathbf{h}_i^TX, \mathbf{m}_j^TY) \\
         &=\max \frac{cov(\mathbf{h}_i^TX, \mathbf{m}_j^TY)}{\sqrt{var(\mathbf{h}_i^TX)var(\mathbf{m}_j^TY)}} \\
         &=\max \frac{\mathbf{h}_i^TC_{XY}\mathbf{m}_j}{\sqrt{(\mathbf{h}_i^TC_{XX}\mathbf{h}_i)(\mathbf{m}_j^TC_{YY}\mathbf{m}_j)}}
    \end{split}
    \end{equation}
where the superscript $T$ of $\mathbf{h}_i^T$ or $\mathbf{m}_j^T$ is the transpose of the vector $\mathbf{h}_i$ or $\mathbf{m}_j$. $C_{XY}$, $C_{XX}$, and $C_{YY}$ are the covariance matrices involving the feature vectors $X$ and $Y$ of the social networks. Vector centralization can avoid the adverse effects of outliers and extreme values. The vectors $X$ and $Y$ obtained in the previous section are zero-centered. Hence, we can calculate their covariance matrices by $C_{XY}=\frac{1}{n}XY^T$, $C_{XX}=\frac{1}{n}XX^T$, and $C_{YY}=\frac{1}{n}YY^T$, respectively.

The objective of Eq. $\ref{rho}$ is equivalent to the following constrained optimization problem since any canonical vector pair $\mathbf{h}_i$ and $\mathbf{m}_j$ are scale-independent.
\begin{equation}\label{CCAobjective}
    \begin{split}
       \max \ & \mathbf{h}_i^TC_{XY} \mathbf{m}_j \\
         s.t. \ & \mathbf{h}_i^TC_{XX}\mathbf{h}_i=1,\ \mathbf{m}_j^TC_{YY}\mathbf{m}_j=1
    \end{split}
    \end{equation}
However, when we have a higher feature dimension $d$ and a relatively smaller number $\mathcal{T}$ of training samples, especially if $d>\mathcal{T}$, the covariance matrices $C_{XX}$ and $C_{YY}$ are singular. The problem is remedied by adding a regularization \cite{Warton2008340} to the covariance matrices as:\\
\begin{equation}\label{CCAreg}
    \begin{split}
       \hat{C}_{XX} &= C_{XX}+r^XI \\
       \hat{C}_{YY} &= C_{YY}+r^YI
    \end{split}
    \end{equation}
where $r^X$  and $r^Y$ are nonnegative regularization coefficients, and $I$ is the identity matrix. The canonical matrices $H$ and $M$ can be obtained by solving a generalized eigenvalue decomposition problem \cite{Hardoon20042639}:\\
\begin{equation}\label{CCAsovle}
    \begin{bmatrix}
      \mathbf{0} & C_{XY} \\
      C_{YX} & \mathbf{0}
    \end{bmatrix}
    \begin{bmatrix}
      H \\
      M
    \end{bmatrix}
    =
    \lambda
    \begin{bmatrix}
      \hat{C}_{XX} & \mathbf{0} \\
      \mathbf{0} &\hat{C}_{YY}
    \end{bmatrix}
    \begin{bmatrix}
      H\\
      M
    \end{bmatrix}
    \end{equation}

In addition, projecting the linked social networks $G^X$ and $G^Y$ to a new common correlated space is possible using the canonical matrices $H$ and $M$. We can link users' identities by calculating the distances of their projected features obtained by $Z^X=H^TX$ and $Z^Y=M^TY$ for $G^X$ and $G^Y$, respectively. This study uses the Euclidean distance as the measurement metric. The distances of users $u_i$ and $u_j$ separately from networks $G^X$ and $G^Y$ in common space $Z$ can be calculated by the square of a binorm:
\begin{equation}\label{Eq.Dist}
    D(z_i^X,z_j^Y )=\|z_i^X-z_j^Y\|_2^2
    \end{equation}
A closer distance $D$ of two users indicates that they have a greater chance of being the same natural person. Consequently, the overall procedure of MAUIL can be given in Algorithm \ref{alg:MAUIL}.

\begin{algorithm}[h!]
    \caption{Overall procedure of MAUIL.}
    \label{alg:MAUIL}
    \begin{spacing}{1.2}
    \begin{small}
    \begin{algorithmic}[1]
    \Require Two social networks $G^X=(V^X,E^X,A^X)$ and $G^Y=(V^Y,E^Y,A^Y)$ , a set of pre-matched user pairs $S$, the dimension $d_c,d_w,d_t$, and $d_s$, the number $k$ of projection vectors.
    \Ensure A set of matched user pairs $L$.
    \State $A^X_c,A^X_w,A^X_t,A^Y_c,A^Y_w,A^Y_t=Preprocess(A^X,A^Y)$ %\Comment{Prepare the  three collections of attribute texts}
    \State $P_c^X / P_c^Y=\Call{CharLevelEmbed}{A^X_c / A^Y_c}$ %\Comment{Learning the multi-level attribute embedding}
    \State $P_w^X / P_w^Y=\Call{WordLevelEmbed}{A^X_w / A^Y_w}$
    \State $P_t^X / P_t^Y=\Call{TopicLevelEmbed}{A^X_t / A^Y_t}$
    \State $P_s^X / P_s^Y=\Call{NetworkEmbed}{E^X / E^Y}$ %\Comment{Learning the network structure embedding}
    \State $X,Y \gets$ Eq. $\ref{NE_comb}$
    \State Standardize each column of $X$ and $Y$ to 0-mean, 1-std.

    \State $X_{train},Y_{train},X_{test},Y_{test}\gets DataSpliting(X,Y,S)$
    \State $H,M=\Call{LearnProjections}{X_{train},Y_{train}}$
    \State $Z^{X}_{test}=H^{T}X_{test}, Z^{Y}_{test}=M^{T}Y_{test}$
    \State $Similarities=EuclideanDistance(Z^X_{test},Z^X_{test})$
    \State $L\gets argmax(Similarities)$
    \State \textbf{return} L
    \Statex
    \Function{LearnProjections}{$X,Y$}
        \State $C_{XX}=\frac{1}{n}XX^T, C_{XY}=\frac{1}{n}XY^T$
        \State $C_{YX}=\frac{1}{n}YX^T, C_{YY}=\frac{1}{n}YY^T$
        \State $\hat{C}_{XX}, \hat{C}_{YY} \gets$ Eq. $\ref{CCAreg}$
        \State $H,M \gets$ Eq. $\ref{CCAsovle}$
        \State \textbf{return} $H, M$
    \EndFunction
    \end{algorithmic}
    \end{small}
    \end{spacing}
	\end{algorithm}

\section{Experiments}
\label{sec:experiment}
This section systematically discusses the datasets, experimental settings and experimental analysis of some significant baseline models. We also perform a sensitivity study on three primary parameters.

\subsection{Datasets}
To evaluate the performance of the MAUIL model, the survey datasets were collected from the internet, including two social networks and two academic coauthor networks.

\begin{figure}[!h]
    \begin{center}
    \includegraphics[width=3.2 in]{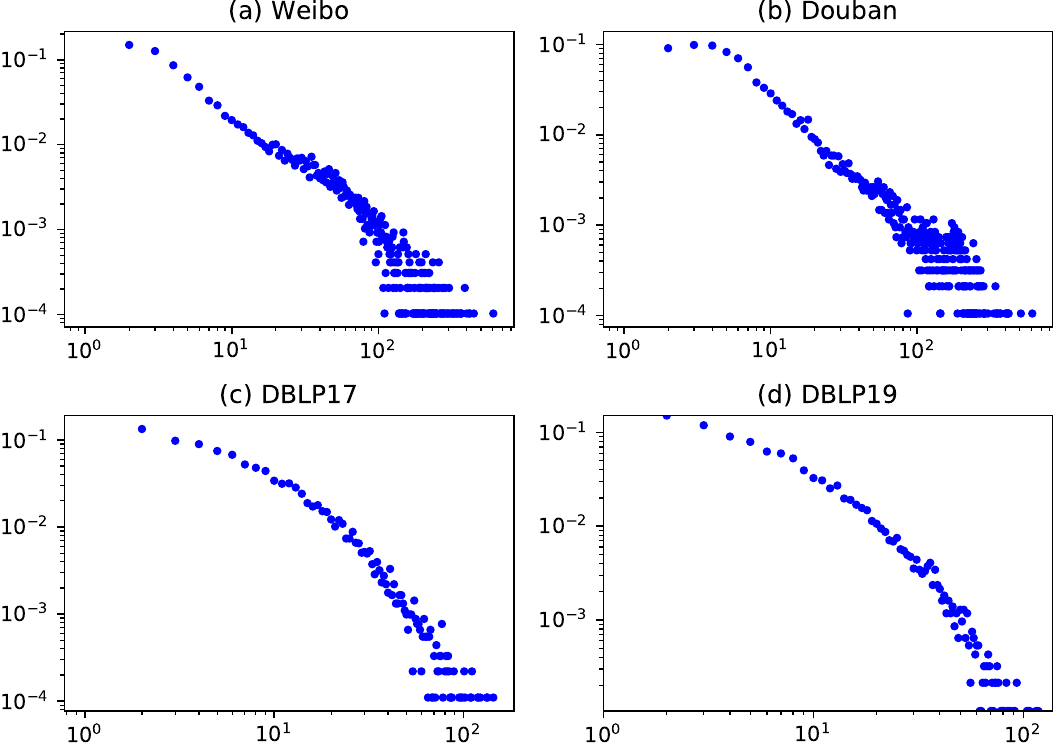}
    \caption{The degree distributions of the networks in the datasets.}\label{Fig:degree}
    \end{center}
    \end{figure}

\textbf{Social networks}: The dataset, denoted Weibo-Douban (WD), refers to two popular Chinese social platforms: Sina Weibo\footnote{https://weibo.com} and Douban\footnote{https://www.douban.com}. We provide an accessible method to build a cross-network dataset WD according to the following steps. First, a small number of users on the Douban platform posted their Weibo accounts on their homepages. These users on the two platforms have real user identity links, which can be used as prealigned user identities in the UIL task. Second, the original data are prepared by crawling users' information pages, including users' attributes and their follower/followee relations. A clear benefit of data crawling in the Weibo platform could not be directly identified in this step. Weibo allows only a small part (currently, two hundred) of follower/followee relations to be returned by crawlers. Hence, the relationships that come from crawlers are quite incomplete. On the other hand, the size of the Weibo network is enormous. Through an empirical process we extract a subnet with common characteristics from the original Weibo network. We repeatedly remove the nodes with degrees less than 2 or more than 1000. Then, the community discovery algorithm \cite{Pares2017229} is performed to find the subnets with typical social network characteristics, including the approximate power-law degree distribution (Figures \ref{Fig:degree}a and \ref{Fig:degree}b) and the high aggregation coefficient. Similar operations are carried out on the Douban network.

\textbf{Coauthor networks}: DBLP is a classic computer science bibliography network. Its data is publicly available\footnote{https://dblp.org/xml/release/}. Each author in DBLP has a unique key, which can be used as the ground truth for the UIL problem. In this study, the DBLP network backups of different periods, i.e.,2017-12-1 and 2018-12-1, were used as target networks to be linked in the UIL experiments. Following previous work \cite{Li2019249}, we select the Turing Award winner \textit{Yoshua Bengio} as the center node in each network, and then delete any nodes more than three steps away from the center.
In addition, the size of two DBLPs is reduced by discovering network communities and repeatedly deleting leaf nodes. The final DBLPs also enjoy the characteristics of a power-law distribution (Figure \ref{Fig:degree}c, \ref{Fig:degree}d) and a high aggregation coefficient. The statistics of the WD and DBLPs are displayed in Table \ref{tab:datasets}.

\begin{table*}[h!]
    \setlength{\tabcolsep}{1pt}
    \renewcommand{\arraystretch}{1.7}
    \caption{The statistics of the datasets used in the experiments.}\label{tab:datasets}
    \begin{footnotesize}
    \begin{center}
    \begin{tabularx}{\textwidth}{p{1.5cm}<{\centering}p{1.4cm}<{\centering}p{1.2cm}<{\centering}p{1.6cm}<{\centering}
                                        p{1cm}<{\centering}p{1cm}<{\centering}p{1cm}<{\centering}p{1cm}<{\centering}p{1.8cm}<{\centering}}
    \toprule[1.0pt]
    \textbf{Datasets}&\textbf{Networks}&\textbf{\#Users}&\textbf{\#Relations}&\textbf{Min. degree}&
    \textbf{Ave. degree}&\textbf{Max. degree}&\textbf{Ave. coeff.}&\textbf{\#Matched pairs}\\
    \hline
    Social		&Weibo    &9,714  &117,218    &2  &12.1  &607    &0.112   &\multirow{2}{*}{1,397}\\
    \cline{2-8}
    networks    &Douban &9,526  &120,245    &2  &12.6  &608    &0.101   &        \\
    \hline
    coauthor	&DBLP17 &9,086  &51,700     &2  &5.7   &144    &0.280   &\multirow{2}{*}{2,832}\\
    \cline{2-8}
    networks	&DBLP19 &9,325  &47,775     &2  &5.1   &138    &0.322   &        \\
    \bottomrule[1.0pt]
    \end{tabularx}
    \end{center}
    \end{footnotesize}
    \setlength{\tabcolsep}{3pt}
    \renewcommand{\arraystretch}{1.2}
	\end{table*}

\subsection{Experimental Settings}
\subsubsection{Baseline Methods}
We select the following baseline methods to assess the comparative performance of the proposed model:
\begin{itemize}
	\item \textbf{IONE}\cite{Liu20161774}: IONE is a semisupervised UIL model that depends on network structures, in which the followership/followeeship of the users are modeled as an input/output context vector. This model sets the weight parameter of the users to 1.

	\item \textbf{PALE-LINE}\cite{Man20161823}: PALE-LINE is a supervised UIL model based on PALE \cite{Man20161823} and LINE \cite{Tang20151067}. A typical operation is embedding the matched networks using the LINE approach. PALE-LINE provides a multilayer perceptron (MLP) that can project the source embeddings to the target space for the UIL task.

    \item \textbf{ABNE}\cite{Liu201923595}: ABNE is a supervised UIL model using graph attention that exploits the social structure by modeling the weighted contribution probabilities between followerships and followeeships.

	\item \textbf{REGAL}\cite{Heimann2018117}: REGAL is an embedding-based unsupervised network alignment model that employs both structures and attributes to capture node similarities. The REGAL experiment in our comparison adopts TF-IDF vectorization and cosine distance measures to address the node attributes and node similarity, respectively.

   	\item \textbf{TADW-MLP}: TADW-MLP is a composite extension of model TADW \cite{Yang20152111} and MLP. TADW is a popular network embedding model that incorporates text attributes under a matrix factorization framework. We apply TADW to embed networks and employ a three-layer MLP that maps the source embeddings to the target space for the UIL problem. In addition, the TF-IDF is also adopted to encode the node attributes.

	\end{itemize}

\textbf{Variants of the proposed MAUIL}: This section also examines three  MAUIL variants to evaluate the effectiveness various model components, including:
\begin{itemize}
	\item \textbf{MAUIL-a}: MAUIL-a ignores the components of the network structure embedding of the MAUIL to assess the component effect of multilevel attribute embedding. Attributes are treated as unique features to address the UIL problem.
	\item \textbf{MAUIL-s}: MAUIL-s attempts to compete with MAUIL-a by retaining only structural features and ignoring attribute features to clarify the component performance of network structure embedding for solving the UIL problem.
	\item \textbf{MAUIL-v}: MAUIL-v directly encodes the feature vectors generated by the MAUIL component of the multilevel attribute embedding, which implements the UIL by comparing the Euclidean distances of the vectors. The model avoids the use of correlated common space and runs in an unsupervised manner.

	\end{itemize}

\subsubsection{Evaluation Metric}
This study evaluates all the comparison methods in terms of Hit-precision \cite{Mu20161775} as well as Precision, Recall and F1-measure. A typical evaluation metric Hit-precision compares the top-$k$ candidates for the identity linkage. The score is computed as follows:\\
\begin{equation}
	h(x)=\frac{k-(hit(x)-1)}{k}
	\end{equation}
where $hit(x)$ is the position of a correctly linked user in the returned list of the top-$k$ candidates. Let $n$ be the number of tested user pairs. The Hit-precision can be calculated using the average score of the successfully matched user pairs: $\frac{1}{n}\sum_{i=1}^{i=n}h(x_i)$. We adopt $k=3$ for all the experimental instances unless otherwise stated.

\subsubsection{Implementation Details}
This study implements the proposed model using the programming language Python 3.7 and runs the codes on a Linux Server with Intel(R) Xeon(R) CPU (E5-2620 v4) and GeForce TITAN X GPU (12 GB memory).

The experiment involved two types of datasets: the social network WD and the coauthor networks DBLPs. First, we explicitly take the network usernames of WD as character-level attributes and convert the Chinese characters into Pinyin characters (Chinese phonetic alphabet). The geographical locations and recent posts of the users are considered word-level and topic-level attributes, respectively. Second, the authors in the DBLPs may have several variants of their names, e.g., \textit{Nicholas Drivalos Matragkas, Nicholas Matragkas}, and \textit{Nikolaos Drivalos}. We randomly choose one of the authors’ names as character-level attributes. Similar to WD, the authors' affiliations in DBLPs are selected as word-level attributes. We pick a random number of articles (at most 100) for each author and take their titles as the topic-level attributes. To illustrate the model's robustness, we add some noise by randomly removing some attribute text items with probability $p=0.2$.

Data preprocessing includes converting all letters of the attributes to lowercase, removing any tonal marks above the letters, eliminating the rare words that occur less than 10 times, excluding stop words such as `the' and `with' and extracting text stems using the NLTK tool \footnote{https://www.nltk.org}. We perform the text segmentation of original texts by adding q-grams with $q=2$ and $q=3$.
In addition, there are several character types in Chinese, such as simplified, traditional, and archaic characters. The experiment requires all Chinese characters to be converted into simplified characters. \footnote{https://pypi.org/project/zhconv/}. Later, the Chinese texts are split into separated words using the tool Jieba \footnote{https://pypi.org/project/jieba/}. The procedure of  word-level embedding includes retaining the neighbor's weight parameter with $\lambda=0.1$. In addition, the word vectors of the Chinese texts are trained based on the Chinese wiki corpus\footnote{https://dumps.wikimedia.org/zhwiki/}. Finally, we set both the hyperparameters $\alpha$ and $\beta$ to $1/d_t$ for the topic-level embedding component of the MAUIL model and its variants.

\subsubsection{Parameter Setup}
This section outlines the model parameters for training MAUIL, including multilevel attribute embeddings, RCCA-based projections and training parameter settings.

\begin{enumerate}[(i)]
\item The dimensions of all features in the multilevel attribute embedding and the structural embedding are the same without loss of generality,
i.e., $D=d_c=d_w=d_t=d_s=100$.
\item The regulation parameters $r^X$ and $r^Y$ for the covariance matrices $C_{XX}$ and $C_{YY}$ are fixed to be $R=r^X=r^Y$. The number of projection vectors is empirically set to $k=25$ and $k=80$ for the WD and DBLPs, respectively. Correspondingly, the regulation parameters $R=10^5$ and $R=10^3$ are considered for the WD and DBLPs, respectively.
\item All experiment instances randomly select $N_{tr}$ and $N_{te}$ matched identity pairs as the training seeds and test data. $N_{tr}$ varies between 100 and 800. $N_{te}$ for the Hit-precision and other metrics (precision, recall and F1) are fixed at 500 and 250, respectively. Another 250 nonmatched user pairs (i.e., negative samples) are randomly generated for precision, recall and F1-measure. The comparative results of all models are evaluated with the training setting $N_{tr}=200$.
\end{enumerate}
Each experimental instance is run 10 times independently. Their average performances are reported and analyzed. The parameters of the comparison models come from the defaults in their papers.

\subsection{Experimental Results}
\subsubsection{Overall Performance}
Tables \ref{tab:comparison1} and \ref{tab:comparison2} summarize the overall results of all the compared methods on the two datasets WD and DBLPs.
The proposed MAUIL method clearly performs better than the comparison baselines on both datasets (at least 15.0\% higher in terms of Hit-precision, 15.49\% higher in F1).

All the comparison methods can be categorized into three groups according to the features that they adopted. The first group performs UIL by using pure network structures, including IONE, PALE-LINE, ABNE, and MAUIL-s. The second leverages only user attributes for solving the UIL problem, i.e., MAUIL-a. The third category combines user attributes and network structures, including REGAL, TADW-MLP, MAUIL-v, and MAUIL. The details of the performance comparison are described below.

First, the experimental results of Group 1 are presented. It is clear from Tables \ref{tab:comparison1} and \ref{tab:comparison2} that the quality of the datasets leads to differences in model performances. The IONE, PALE-LINE, ABNE, and MAUIL-s methods perform better on DBLPs than on WD. For example, INOE registers a higher Hit-precision ($k=3$) by 9.2\% (10.51\% in terms of F1), PALE-LINE by 10.5\% (15.67\% in F1), ABNE by 11.4\% (14.96\% in F1), and MAUIL-s by 16.4\% (19.18\% in F1) on DBLPs. This is mainly because of the lower data structure integrity in the WD real social networks, which cannot provide valid clues for recognizing user identities. However, the four models' most prominent highest score is of the proposed variant MAUIL-s that registers at least 5.6\% (6.43\% in F1) higher than the others on DBLPs.
Following the performance scores of PALE-LINE and MAUIL-s, a higher score of MAUIL-s was recorded, although both models depend on the LINE structure embedding method. As a result, MAUIL-s performs 4.7\% and 10.6\% better than PALE-LINE on the WD and DBLPs, respectively. MAUIL and its variants provide a linear projection that maps the two networks to the common space. The precision advantages of MAUIL-s prove that the established mapping on the MAUIL model series does indeed benefit the UIL task since MAUIL and its variants possess the same model component to deal with network structure embedding.

\begin{table}[!htb]
    \setlength{\tabcolsep}{1pt}
    \caption{Comparison with the baseline methods in terms of Hit-precision score.}\label{tab:comparison1}
    \begin{footnotesize}
    \begin{center}
    \renewcommand{\arraystretch}{1.5}
    \begin{tabularx}{8.5cm}{p{2cm}<{\centering}|
                                    p{1.0cm}<{\centering}p{1.0cm}<{\centering}p{1.0cm}<{\centering}|
                                    p{1.0cm}<{\centering}p{1.0cm}<{\centering}p{1.0cm}<{\centering}}
    \toprule[1.2pt]
    \multirow{2}{*}{Method}&\multicolumn{3}{c}{Weibo-Douban} &\multicolumn{3}{c}{DBLP17-DBLP19} \\
    \cline{2-7}
    			&	k=1		&	k=3		&	k=5		&	k=1		&	k=3		&	k=5		\\
    \hline
    IONE   		&	0.013 & 0.022&  0.031	&	0.076 	&  0.114 	&    0.140\\
    PALE-LINE 	&	0.013 & 0.025&  0.036	&	0.085   &  0.130    &    0.165\\
    ABNE        &	0.049 & 0.066&	0.077	&  0.133 	&  0.180 	&    0.211\\
    REGAL      	&	0.016 &	0.030& 	0.041   &  0.411 	&  0.436 	&    0.452\\
    %TADW-MLP    &	0.004 & 0.007&	0.011	&  0.188 	&  0.261 	&    0.313\\
    ~TADW-MLP&~0.004&~0.007&~0.010&~0.380&~0.504&~0.575\\
    \hline
    MAUIL-v		&	0.238      &	0.280     &	    0.306	 &	   0.628    &   0.655      &   0.695      \\
    MAUIL-s		&	0.044      &    0.072     &     0.095    &     0.159 	&   0.236 	   &   0.289      \\
    MAUIL-a	    &\textbf{0.313}&    0.368     &     0.400    &\textbf{0.661}&   0.690      &   0.706      \\
    MAUIL       &\textbf{0.313}&\textbf{0.373}&\textbf{0.408}&      0.660   &\textbf{0.702}&\textbf{0.725}\\
    \bottomrule[1.2pt]
    \end{tabularx}
    \end{center}
    \end{footnotesize}
    \setlength{\tabcolsep}{3pt}
    \renewcommand{\arraystretch}{1.2}
	\end{table}

\begin{table}[!htb]
    \setlength{\tabcolsep}{1pt}
    \caption{Comparison with the baseline methods in terms of precision, recall and F1-measure (\%).}\label{tab:comparison2}
    \begin{footnotesize}
    \begin{center}
    \renewcommand{\arraystretch}{1.5}
    \begin{tabularx}{8.5cm}{p{2cm}<{\centering}|
                                    p{1.0cm}<{\centering}p{1.0cm}<{\centering}p{1.0cm}<{\centering}|
                                    p{1.0cm}<{\centering}p{1.0cm}<{\centering}p{1.0cm}<{\centering}}
    \toprule[1.2pt]
    \multirow{2}{*}{Method}&\multicolumn{3}{c}{Weibo-Douban} &\multicolumn{3}{c}{DBLP17-DBLP19} \\
    \cline{2-7}
    			&	Pre.	&	Rec.	&	F1		&	Pre.	&	Rec.	&	F1		\\
    \hline
    IONE   		&	1.20      & 70.17     &  2.36	&  7.00   &  82.2 	&    12.87\\
    PALE-LINE 	&	1.80      & 85.37     &  3.52	&  10.76  &  89.87  &    19.19\\
    ABNE        &	4.08      & 76.14     &  7.73	&  13.08  &  87.16 	&    22.69\\
    REGAL      	&	1.24      & 95.25     &  2.44   &  32.85  &  80.01 	&    46.54\\
    TADW-MLP    &	0.35      & 48.87     &  0.69	&  30.07  &  75.12 	&    42.91\\
    \hline
    MAUIL-v		&	23.05       &\textbf{95.34}&	37.09      &	48.07      &	75.87      &   58.82      \\
    MAUIL-s		&	5.26        &      94.63   &    9.94       &    17.19      &\textbf{96.15} &   29.12      \\
    MAUIL-a     &   27.68       &      88.69   &    42.15      &    50.01      &    75.50      &   60.12      \\
    MAUIL       &\textbf{28.12} &      89.37   &\textbf{42.72} &\textbf{51.63} &    77.77      &\textbf{62.03}\\
    \bottomrule[1.2pt]
    \end{tabularx}
    \end{center}
   Note: The Hit-precision scores are evaluated in 500 matched test pairs, while precision, recall and F1 scores in 250 matched test pairs plus 250 non-matched user pairs.
    \end{footnotesize}
    \setlength{\tabcolsep}{3pt}
    \renewcommand{\arraystretch}{1.2}
	\end{table}

Second, the experimental results of Group 2 involving MAUIL-a are exhibited, which are only concerned with user attributes. MAUIL-a achieves Hit-precision scores of 0.368 and 0.69 on WD and DBLPs, respectively. The more adequate and cleaner user attribute information in DBLPs compared with WD creates a significant 0.322 difference in model precision. Users’ blogs and posts on WD are normally inconstant and vary greatly in content, number of posts, and publishing frequency. What stands out in Table \ref{tab:comparison1} is that MAUIL-a performs significantly better than the structure-based models (in Group 1) by at least 29.6\% on WD and at least 45.4\% on DBLPs in terms of Hit-precision. The abundant attribute information in our datasets has an apparent positive influence on the model's effect.

Third, Group 3 incorporates both attribute and structure features to investigate the UIL task. Both REGAL and TADW-MLP can extract attribute features. However, they do not distinguish attribute text types or high-level semantic features. A comparison of the scores among those models in Group 3 confirms that the utilization and differentiation of multitype user attributes are beneficial improving a model's performance. For example, MAUIL-v performs considerably better than REGAL by 25\% on WD in terms of Hit-precision, and also recorded a better score than TADW-MLP by 27.3\%.

The character-level attributes (user names) of social networks are the strongest among all attribute types, while the word-level (locations) and topic-level (posts) attributes are more diverse and noisy. Consequently, character-level attributes a dominate model's performance.

\begin{figure}[!h]
    \begin{center}
    \includegraphics[width=3.0 in]{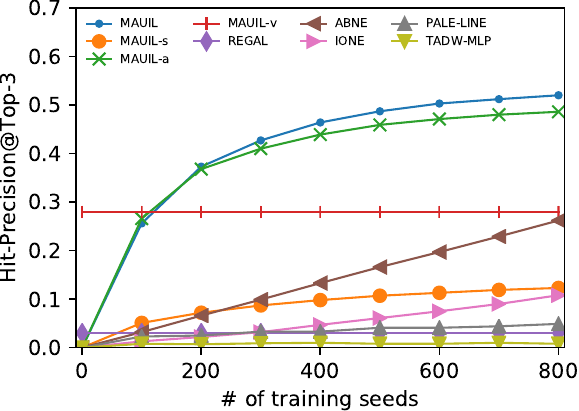}
    \caption{Hit-precision performance on WD w.r.t the increase of training seeds $N_{tr}$.}\label{Fig:Seeds_wd}
    \end{center}
    \end{figure}

\subsubsection{The effect of different training seeds}
Performance trend tests were used to analyze the comparison models' Hit-precision with the increase in training seed parameters $N_{tr}$.

First, the test results of all the comparison models in dataset WD are shown in Figure \ref{Fig:Seeds_wd}. Real social networks are often sparse in structure. Hence, the performance improvement of the simple structure methods, i.e., IONE, PALE-LINE, ABNE, and MAUIL-s, is very limited. Figure \ref{Fig:Seeds_wd} provides an example of the above statement. The test scores of the pure structure-based methods on WD rise slowly by increasing $N_{tr}$ from 100 to 800. The comparison of their scores with the proposed MAUIL confirms that combining network structures and user attributes causes the scores to reach a peak and has an advantage of at least 25.8\% over all structural baselines when $N_{tr}=800$.

\begin{figure}[!h]
    \begin{center}
    \includegraphics[width=3.0 in]{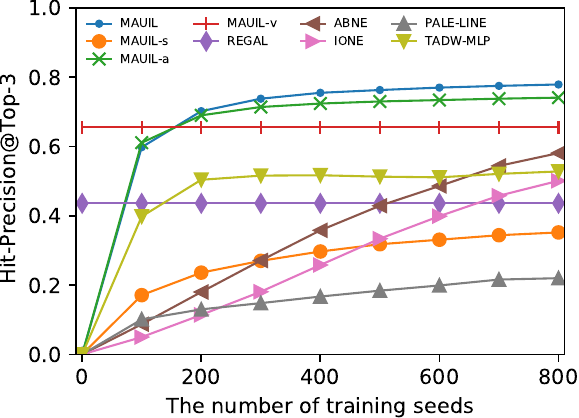}
    \caption{Hit-precision performance on DBLPs w.r.t the increase of training seeds $N_{tr}$.}\label{Fig:Seeds_dblp}
    \end{center}
    \end{figure}

In addition, we consider a similar test on DBLP datasets. The result is shown in Figure \ref{Fig:Seeds_dblp}.
The performance of all methods except for REGAL and MAUIL-v present similar upward trend curves as $N_{tr}$ increases from 100 to 800. For instance, IONE's performance grows from 0.05\% to 50.1\% on DBLPs in terms of Hit-precision and on ABNE the performance grows from 0.087\% to 58.1\%. Consequently, more training data contributed to the precision increase. Noticeably, MAUIL achieves the best scores across all $N_{tr}$ settings.
Under the semisupervised setting, MAUIL can effectively leverage limited annotations and rich user features to ensure its superiority. Therefore, MAUIL is expected to be well adapted to annotate constrained scenarios.

Finally, what can be clearly seen in Figures \ref{Fig:Seeds_wd} and \ref{Fig:Seeds_dblp} are the straight red lines referring to the Hit-precision of MAUIL-v with the seed increase on both datasets. The unsupervised working mode of MAUIL-v is responsible for this stable performance.

\begin{figure}[!h]
    \begin{center}
    \includegraphics[width=3.2 in]{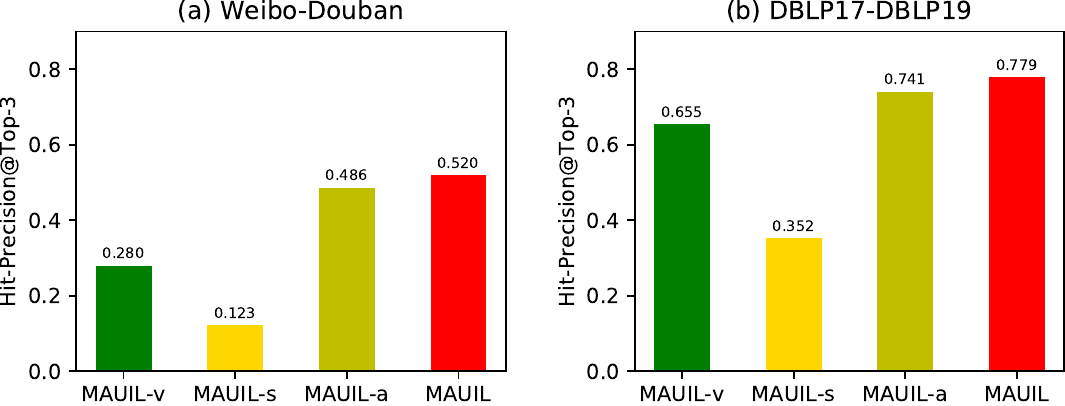}
    \caption{Hit-precision performance of MAUIL and its variants ($N_{tr}=800$).}\label{Fig:components}
    \end{center}
    \end{figure}

\subsubsection{The contribution of different components of MAUIL}
As seen from Figures \ref{Fig:Seeds_wd} and \ref{Fig:Seeds_dblp}, MAUIL always outperforms its three variants, i.e., MAUIL-s, MAUIL-a, and MAUIL-v, across all training seeds $N_{tr}$. This section uses $N_{tr}=800$ as an example to assess the contributions of each MAUIL component.

The columnar analysis shown in Figure \ref{Fig:components} illustrates that MAUIL achieces 24\% higher Hit-precision scores than MAUIL-v on WD and 12.4\% higher on DBLP. The priorities demonstrate the power of MAUIL when the model adds the step of projecting the linked networks into the common correlated space.

On the other hand, MAUIL performs better than MAUIL-s by 39.7\% in terms of Hit-precision on WD and 42.7\% on DBLP. The advantages are caused by incorporating attribute information. In addition, MAUIL also beats MAUIL-a by 3.4\% on WD and 3.8\% on DBLP. Utilizing network structural information indeed benefits the UIL tasks. However, the role of the network structures does not have as strong an impact as that of the attribute information in our datasets. Thus, MAUIL is quite impressive when attribute information is sufficient.

\subsubsection{Parametric Sensitivity Analysis}
This section investigates the parameter sensitivity of the proposed MAUIL on three primary parameters: (1) the feature dimensions $D$, (2) the number of projection vectors $K$, and (3) the regularization coefficient $R$.

\begin{figure}[!h]
    \begin{center}
    \includegraphics[width=3.2 in]{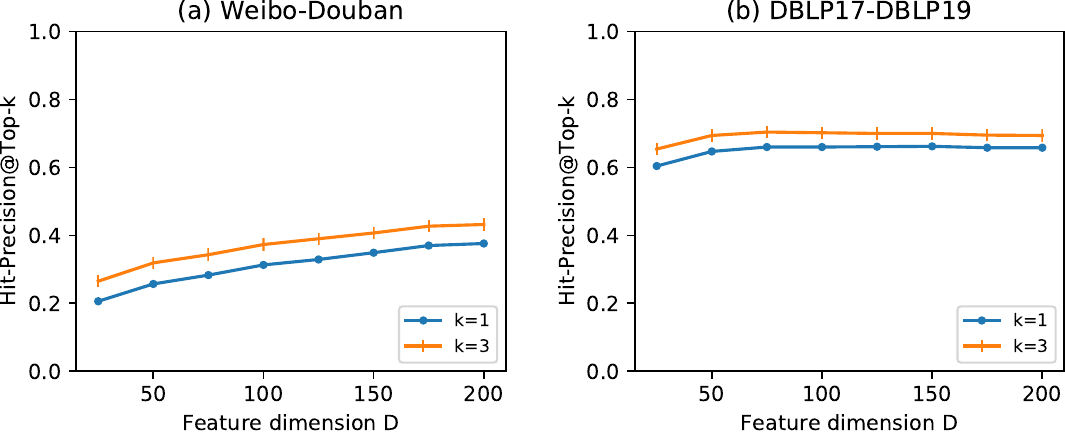}
    \caption{Hit-precision performance w.r.t the increase of the feature dimension $D$.}\label{Fig:dim}
    \end{center}
    \end{figure}

Figure \ref{Fig:dim} portrays the Hit-precision variation in regards to the value change of feature dimension parameter $D$. We can observe that the MAUIL performance scores increase at the beginning and then remain stable as the dimension rises on both datasets. This trend means that a more top dimensional feature space helps to better preserve the user information, which further contributes to the performance improvement of the UIL models.
In the actual operation process, too high of a dimension will increase the consumption of the computing resources. Hence, we select $D=100$ to a trade-off between effectiveness and efficiency.

\begin{figure}[!h]
    \begin{center}
    \includegraphics[width=3.2 in]{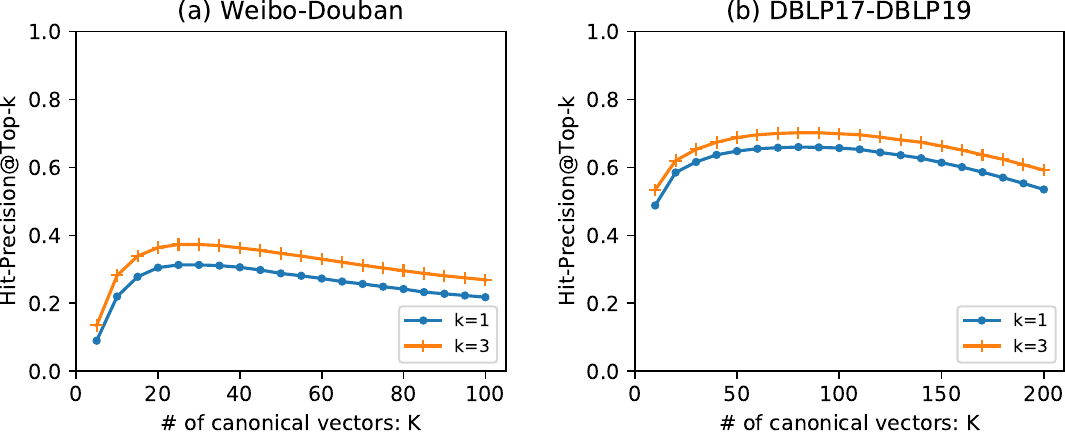}
    \caption{Hit-precision performance w.r.t the increase pf projection vectors $k$.}\label{Fig:K}
    \end{center}
    \end{figure}

Figure \ref{Fig:K} depicts the sensitivity of the models' performances when the value of the projection vector $k$ changes. MAUIL presents a similar trend in both of the datasets. Its performance becomes optimal when $k$ fluctuates from 25 to 35 on WD and from 60 to 110 on DBLP. Therefore, if $k$ changes within a reasonable range, MAUIL will remain stable.

\begin{figure}[!h]
    \begin{center}
    \includegraphics[width=3.2 in]{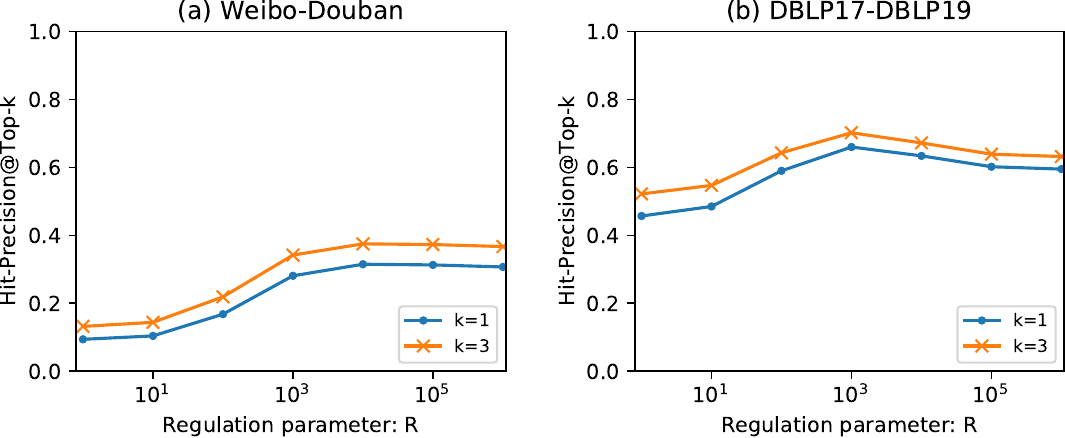}
    \caption{Hit-precision performance w.r.t the increase of regulation parameter $R$.}\label{Fig:reg}
    \end{center}
    \end{figure}

Figure \ref{Fig:reg} illustrates the performance with different settings of the regulation parameter $R$. We set the range from $0$ to $10^6$ for $R$ across the two datasets. The Hit-precision increases gradually and remains steady after $R>10^4$ on WD. However, the Hit-precision slightly drops after $R$ reaches $10^3$ on DBLP. One possible reason is that the parameter's variation is exponential, and the stable interval of $R$ is slightly smaller than that of the exponential change. Hence, it is necessary to choose $R$ in a small range around the $10^3$ locus.

\section{Conclusion}
\label{sec:conclusion}
The purpose of the present study was to investigate the problem of user identity linkage between multiple social networks. A high-precision method for user identity discovery can significantly improve the reliability of AI applications such as recommender systems, search engines, information diffusion predictions, cyber identity and criminal stalking. Hence, this work has constructed a novel semisupervised model, namely, MAUIL, to seek the potential user identity between two attributed social networks. Semisupervision insight greatly reduces the model's dependence on data annotation. Compared with other models, MAUIL synthesizes three kinds of user attributes, i.e., character-level, word-level and topic-level attributes and network structure features, to obtain the best Hit-precision scores for the UIL problem. The experimental analysis shows that the combination of different text features in social networks can significantly improve the resolution accuracy of the problem. Additionally, the RCCA-based linear projections designed in MAUIL revealed that maximizing the correlations between available feature vectors from different views has benefits in finding the same natural persons across multiple social networks.

This study extensively evaluated the proposed model on two real-world datasets of social networks and coauthor networks. The results show the superior performance of MAUIL by a comprehensive comparison with state-of-the-art baseline methods. A limitation of this study is that the dataset WD in this paper is merely a subset of the current social networks with similar structure distributions. We don't have the privacy rights to obtain the full data. Finally, this paper complimentarily provides the complete program code and datasets. We highly recommend the suggested UIL problem-solving framework and technical update directions for other interested researchers.

\end{document}